\documentstyle[mypp,epsf,twoside]{article}

\def\xis{\xi_{_{S}}}
\def\xir{\xi_{_{R}}}
\def\hmpc{\mbox{$h^{-1}$~\rm{Mpc}}}
\def\kms{\mbox{\rm km/s}}
\def\disp{{_{\! D}}}
\def\linflow{{{\! _L}}}
\def\pr{P_{_{\! R}}}
\def\ps{P_{_{\! S}}}

\def\sigv{\sigma_{v}}
\def\iras{\mbox{IRAS 1.2Jy}}

\def\gsim{\lower 2pt \hbox{$\, \buildrel {\scriptstyle >}\over {\scriptstyle
\sim}\,$}}
\def\lsim{\lower 2pt \hbox{$\, \buildrel {\scriptstyle <}\over {\scriptstyle
\sim}\,$}}

\newcommand{\beq}{\begin{equation}}
\newcommand{\eeq}{\end{equation}}
\newcommand{\comment}[1]{}

\newcommand{\myvec}[1]{{\bf #1}}
\newcommand{\myunitvec}[1]{{\bf #1}}

\begin{document}

\markboth{Bromley, Warren \& Zurek}%
{Estimating $\Omega$ from Galaxy Redshifts}
\pagestyle{myheadings}
\thispagestyle{empty}

\title{
 Estimating $\Omega$ from Galaxy Redshifts: 
 Linear Flow Distortions and Nonlinear Clustering
}

\author{
 B. C. Bromley\altaffilmark{1,2}, 
 M. S. Warren\altaffilmark{1}, and W. H. Zurek\altaffilmark{1}
}

\affil{
\parbox{5.75in}{$^1$Theoretical Astrophysics, MS B288, 
Los Alamos National Laboratory,
Los Alamos, NM 87545
\\
$^2$Theoretical Astrophysics, MS-51, Harvard-Smithsonian
Center for Astrophysics, 60 Garden Street, Cambridge, MA 02138}}

\begin{abstract}

We propose a method to determine the cosmic mass density $\Omega$ from
redshift-space distortions induced by large-scale flows in the
presence of nonlinear clustering.  Nonlinear structures in redshift
space such as fingers of God can contaminate distortions from linear
flows on scales as large as several times the small-scale pairwise
velocity dispersion $\sigv$.  Following Peacock \& Dodds (1994), we
work in the Fourier domain and propose a model to describe the
anisotropy in the redshift-space power spectrum; tests with
high-resolution numerical data demonstrate that the model is robust
for both mass and biased galaxy halos on translinear scales and above.
On the basis of this model, we propose an estimator of the linear
growth parameter $\beta = \Omega^{0.6}/b$, where $b$ measures bias,
derived from sampling functions which are tuned to eliminate
distortions from nonlinear clustering.  The measure is tested on the
numerical data and found to recover the true value of $\beta$ to
within $\sim 10$\%.  An analysis of \iras\ galaxies yields $\beta =
0.8^{+0.4}_{-0.3}$ at a scale of 1,000~\kms\ which is close to optimal
given the shot noise and the finite survey volume.  This measurement
is consistent with dynamical estimates of $\beta$ derived from both
real-space and redshift-space information.
The importance of the method presented here is that nonlinear
clustering effects are removed to enable linear correlation anisotropy
measurements on scales approaching the translinear regime.  We discuss
implications for analyses of forthcoming optical redshift surveys in
which the dispersion is more than a factor of two greater than in the
IRAS data.


\end{abstract}

\section{Introduction}

A redshift-space map of galaxies is distorted relative to the
real-space galaxy distribution as a result of peculiar motions along
an observer's line of sight.  These distortions generate an anisotropy
in pairwise correlations which would not appear in the real-space
galaxy distribution of a statistically homogeneous and isotropic
universe.  Kaiser (1987) quantified the correlation anisotropy that
results from large-scale peculiar flows in terms of the power spectrum
of galaxies using the linear theory of gravitational instability.  He
demonstrated that in the linear regime the power measured by a distant
observer only depends on the angle between the wavevector and the
observer's line of sight, and on the dimensionless factor $\beta \equiv
\Omega^{0.6}/b$.  Here, $\Omega$ is the cosmic mass density parameter
and $b$ is a function that differs from unity if the galaxies are a
biased sample of the total mass.  Hamilton (1992) transformed Kaiser's
result out of the Fourier domain to determine the redshift-space
correlation function $\xis$ in linear theory and proposed an estimator
of $\beta$ based on a spherical harmonic decomposition of
$\xis$. Subsequently, a number of $\beta$ measurements from linear
flow distortions have been reported (e.g., Hamilton 1993, 1995;
Bromley 1994; Fisher, Scharf \& Lahav 1993; Fisher et al. 1994; and
Nusser \& Davis 1994).

There are two major challenges for a statistical measurement of
$\beta$ from linear flows in a real galaxy redshift catalog. The first
comes from the finite size of the catalog. The catalog must contain a
fair sample of structure on a particular scale or else the correlation
information will contain finite-sampling noise.  A reasonable
criterion is that sampling scales should be below $\sim 10$\% of the
characteristic size of survey volume.  The second challenge is to
ensure that the sampling scales are large enough so that signal from
nonlinear clustering does not contaminate the linear fluctuation
modes.  Fingers of God can extend up to several thousand kilometers
per second in optically selected galaxy surveys and clearly affect the
redshift-space power on these scales.
(Here redshift-space distances are usually given in terms \kms; where
real-space lengths are needed, we use the conventional units of
\hmpc\ where $h$ is the Hubble parameter in units of 100~\kms.)
Thus the signal in redshift space from linearly growing
fluctuation modes on scales from $\sim$1,000~\kms\ (characteristic
of the translinear regime below which linear theory breaks down)
to $\sim$4,000~\kms\ can be contaminated by strongly nonlinear
features.  Present-day catalogs are typically within a few times
10,000~\kms\ in radial extent, leaving at best only a small range of
scales on which purely linear modes can be measured with a high degree
of statistical integrity.

In this paper we address this problem of determining $\beta$ from
linear flows in the presence of nonlinear clustering.  Our approach is
motivated by a remarkable result from Peacock \& Dodds (1994)
concerning the redshift-space power spectrum $\ps$.
While $\xis$ itself can be modeled given the distribution of
pairwise velocities, there is evidence (Fisher et al. 1994; Warren 1995)
that this distribution has complicated behavior on a critical
range scales from $\sim$1--10~\hmpc.
In contrast, Peacock \& Dodds (1994) found that nonlinear signal can
be modeled simply and accurately in the Fourier domain on translinear
scales and above. Specifically, the effects nonlinear clustering and
linear flows in redshift space can be described by separable linear
filters acting on the real-space power spectrum.  The linear filtering
hypothesis has been supported by tests with $N$-body simulations,
although only mass particles and not collapsed galaxy halos were
considered (Gramann, Cen \& Bahcall 1993; Tadros \& Efstathiou 1996).
Here, we first suggest a form of the filters, which amounts to
constructing a model for the anisotropy in $\ps$, and determine its
validity in high-resolution numerical simulations using both mass
particles and collapsed halos which are identified as galaxies. On the
basis of this model we then suggest a method to extract $\beta$ on
scales of 1,000~\kms\ and above, even when the velocity dispersion is
of comparable size.  Finally, we apply our method to the \iras\
survey. The result is a new measure of $\beta$ on translinear scales.

\section{Power in Redshift Space}

Following Peacock \& Dodds (1994) we assume that the effects of
line-of-sight peculiar motion on a galaxy distribution can be modeled
by a linear filter function $F$ that relates the real-space and
redshift-space power spectra:
\beq\label{eq:pfilt}
\ps(k,\mu) = F(k,\mu) \pr(k) \ ,
\eeq
where $\mu$ is the angle cosine of the observer's line of sight and
a wavevector of amplitude $k$. For definiteness we assume that the
power is measured in a distant compact survey so that $\mu$ is
constant for all points in the survey.

The form of $F$ can be motivated from its limiting behavior at small
and large scales. At small scales, the peculiar motions are dominated
by isotropic dispersion, and the redshift-space correlation function
obtains from a simple convolution of the real-space function $\xir$
and the distribution function $f_v$ of pairwise radial velocities.
According to evidence from simulations (Zurek et al. 1994) and
observations (Marzke et al. 1995), $f_v$ is accurately described by an
isotropic exponential function at small pair separation,
\beq\label{eq:vdist} 
f_v \sim \exp(-\sqrt{2}|v|/\sigma_v) \ ,
\eeq
where $\sigma_v$ is the pairwise radial velocity dispersion.  In the
Fourier domain, the redshift-space power spectrum is thus related to
the real-space power spectrum by a linear filter, $F_\disp$, the
Fourier transform of $f_v$ taken as a function of line-of-sight
pairwise velocity:
\beq\label{eq:pfilt_disp}
F_\disp(k,\mu) = \left( 1 + \frac{\mu^2 k^2 \sigma_v^2}{2} \right)^{-1} .
\eeq

Turning to the large-scale regime, Kaiser's (1987) formula for
a filter $F_\linflow$ which accounts for the anisotropy in redshift-space
power from linear flows is
\beq\label{eq:pfilt_linflow}
F_\linflow = (1+\beta \mu^2)^2 \ .
\eeq

Peacock \& Dodds (1994) found that an effective model for the desired
filter $F$ is obtained from a product of $F_\disp$ and
$F_\linflow$. Following their ansatz we write
equation~(\ref{eq:pfilt}) as
\beq\label{eq:pfilt_explicit}
\ps(k,\mu) = \frac{(1+\beta \mu^2)^2}{1+\sigma_v^2\mu^2 k^2/2} \pr(k) \ .
\eeq
In committing to this model (eq.~[\ref{eq:pfilt_explicit}]) we limit
our analysis to scales above the translinear regime.  There is
evidence (Brainerd et al.~1996) that Kaiser's formula
(eq.~[\ref{eq:pfilt_disp}]) for the anisotropy in the redshift-space
power spectrum breaks down at translinear scales. This breakdown can
be quantified by including a ``softening parameter'' of roughly
translinear scale into equation~(\ref{eq:pfilt_disp}) which drives the
filter to unity in the strongly nonlinear regime (see equation [5] in
Brainerd et al.~1996). Theoretical insight into this effect is given
by Fisher \& Nusser (1996).

The model in equation (\ref{eq:pfilt_explicit}) also may be limited by
the assumption that the small-scale velocities are random and
isotropic and exhibit none of the coherent flows that are generated by
the linear or quasilinear dynamics.  Numerical simulations support
this assumption on megaparsec scales and below.  A naive linear theory
extrapolation of the pairwise velocity dispersion by way of the
cosmological pair conservation equation (Peebles 1980 \S71) also
suggests that the linear flow contribution to $\sigma_v$ at 1~\hmpc\
is small for most plausible cosmological scenarios.  In the cold dark
matter model discussed below, linear flows would contribute (in
quadrature) only $\sim 100$~\kms\ to the total of $\sigma_v \approx
1,100$, which amounts to less than a percent in $\sigma_v^2$.  However
the situation is more complicated on translinear scales where both
random dispersion and significant coherent flows exist. Fortunately,
the random motion contribution to the anisotropy of redshift-space
fluctuation modes above the translinear regime comes primarily from
strongly nonlinear scales where the velocity dispersion and spatial
pairwise correlations are strongest and where the velocity
distribution function takes a particularly simple form.

We note that the above expression differs from the result of Peacock
\& Dodds in the choice of the model for the small-scale velocity
distribution function. Their work was based on a Gaussian model,
however, as they and others (Cole, Fisher \& Weinberg 1995; Brainerd
et al. 1996) have noted, the difference between a Gaussian
distribution and the exponential distribution is of little consequence
in the quantification of redshift-space distortions.  Nonetheless we
advocate the exponential model because of its successes in describing
both real and simulated data.

To determine the efficacy of the parameterization in
equation~(\ref{eq:pfilt_explicit}) we examine the redshift-space power
in numerical simulations. We ran two 17 million particle
high-resolution N-body simulations in periodic cubes of size
125~\hmpc\ and 500~\hmpc, respectively, using a treecode with force
smoothing of 0.01~\hmpc. In both cases the primordial power spectrum
was based on cold dark matter (CDM) normalized to $\sigma_8 = 0.74$ in
an $\Omega = 1$, $h = 0.5$ universe. This cosmogony provides a
reasonable match to the observed spatial clustering and pairwise
velocity statistics of optically selected galaxies (Zurek et al. 1994;
Marzke et al. 1995; Brainerd et al. 1996). We use the large-volume
simulation to study the redshift distribution of mass on scales
greater than 50~\hmpc; the small-volume run has sufficient mass
resolution to identify halos as realistic galaxy candidates, hence it
is used to study both the mass and galaxy distributions on smaller
scales.

The results of the N-body analysis are presented in Figure~1, showing
the anisotropy in redshift-space power spectrum of the CDM mass
(Fig. 1a) and the full catalog of halos (Fig. 1b).  We plot the power
at constant $k$ values as a function of the angle between the
wavevector and the observer. Using the megaparsec separation value of
$\sigv$, with values of 1,100~\kms\ and 850~\kms\ for mass and halos
respectively, we find that the parameterization in
equation~(\ref{eq:pfilt_explicit}) is successful on scales near and
above the translinear regime.  The breakdown in the model occurs on
scales which are below $\sigv$, corresponding roughly to the
characteristic size of the Fingers of God. 

\begin{figure}[p]
 \centerline{
  \epsfxsize=3.0in\epsfbox{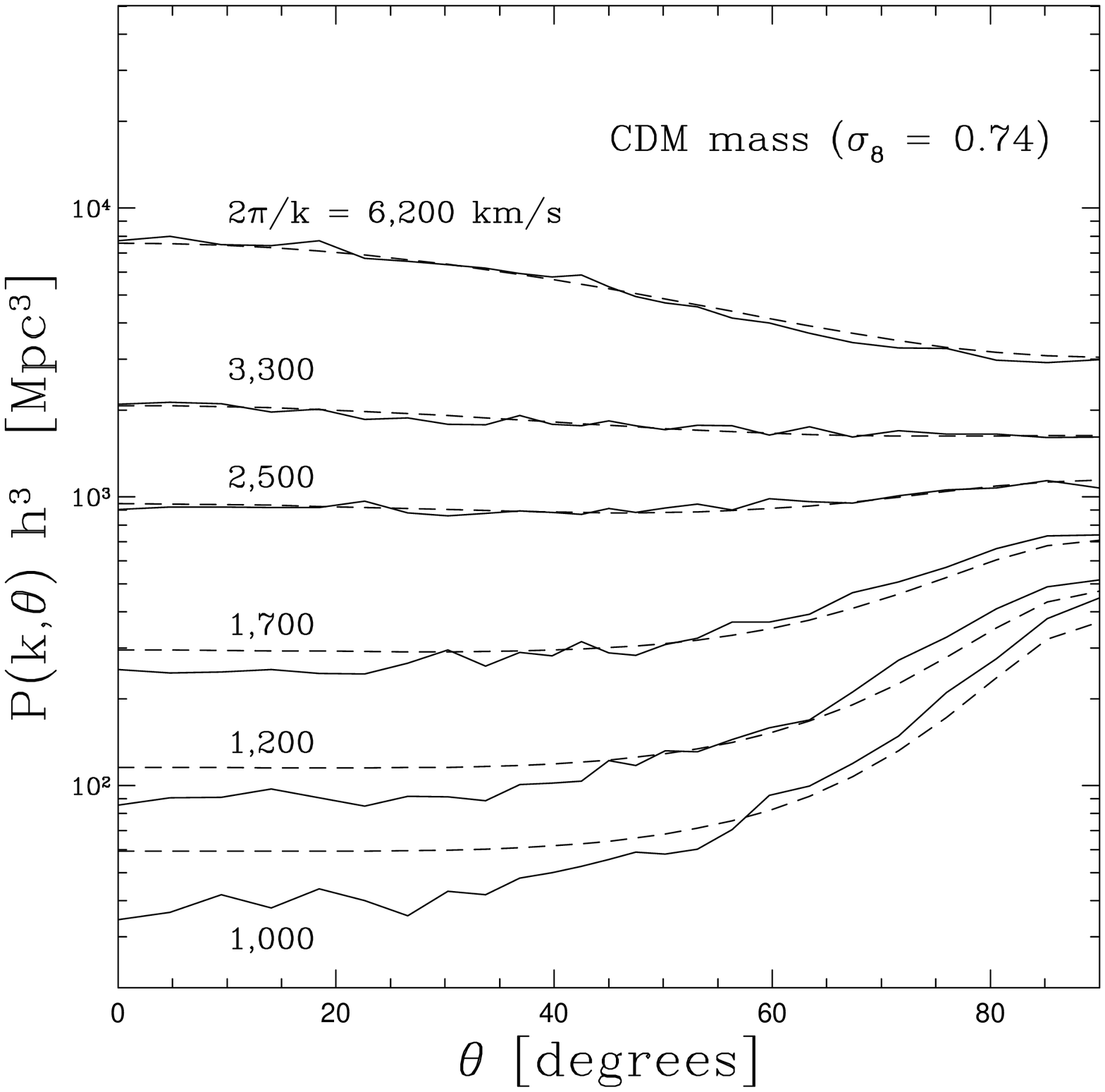}
}
\centerline{
  \epsfxsize=3.0in\epsfbox{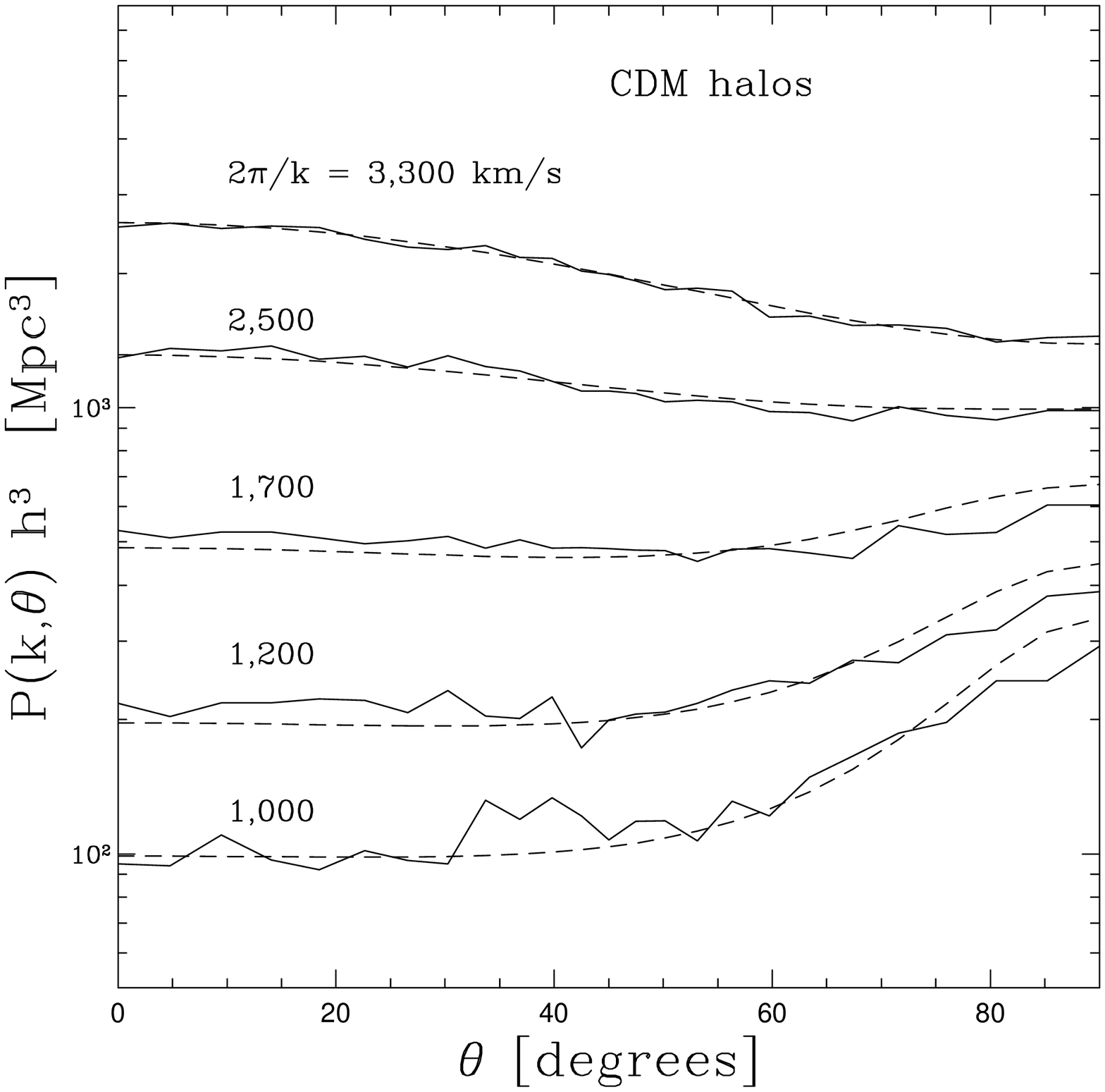}
  \epsfxsize=3.0in\epsfbox{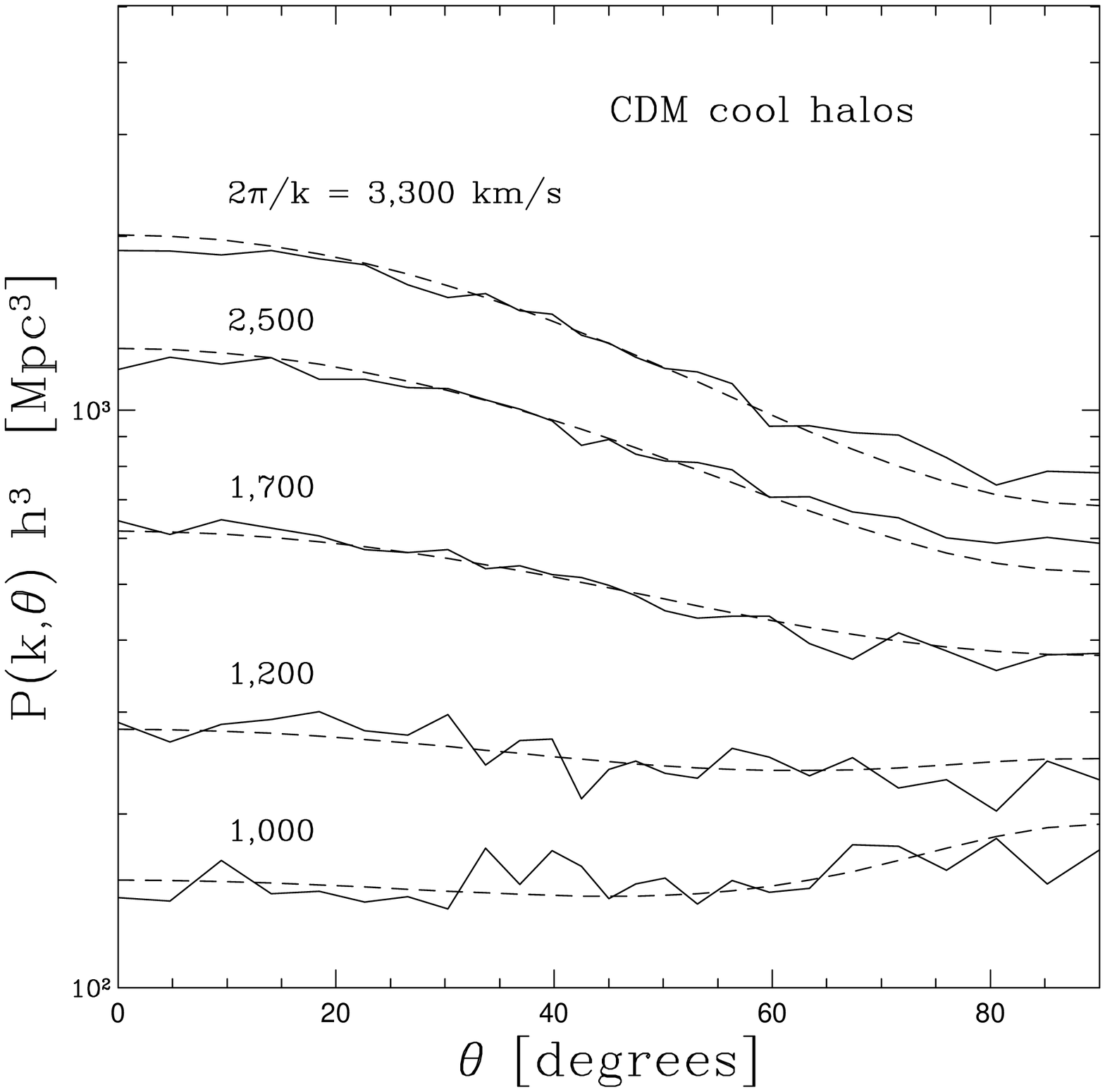}
}
\caption{\protect\small
The two-dimensional redshift-space power spectrum.
The power is shown as a function of the angle between the observer's
line of sight and the wavevector for several different wave magnitudes
$k$. The plots show the power for (a) the mass distribution, (b) the
galaxy halos, and (c) the culled halo subset with low small-scale
velocity dispersion.  The solid lines show the estimate of $P$ from
the simulation data, while the dashed curves are the predicted power
using the multiplicative filter model in equation~(5). A $\beta$ value
of unity was used to generate the model curves in all three plots,
while the $\sigv$ values were 1,100~\kms, 850~\kms, and 400~\kms, for
the mass, halos, and cool halos, respectively.
}
\end{figure}

Remarkable is the extent to which the dispersion signal affects the
angular dependence of the redshift-space power. In the case of the
CDM mass distribution, there is negligible anisotropy at $2\pi/k =
3,300$~km/s, three times $\sigv$, because the large-scale enhancement
in power along the line-of-sight is almost completely canceled by
small-scale dispersion.

Figure 1c depicts results from a third dataset, a cool subset of halos
obtained by eliminating all objects which are within 3 Abell radii of
the highest density peaks and which have peculiar velocities greater
than 750~\kms.  Although produced in a rather contrived manner, this
culled dataset is suggestive of the \iras\ galaxies in that they both
have low velocity dispersions compared with optical galaxies and have
relatively small numbers of objects in the cores of rich optical
clusters.  Indeed, the resulting catalog has $\sigma_v$ of 384~\kms\
and spatial clustering properties below 100~\hmpc\ which are
consistent with the \iras\ data (Brainerd et al. 1996). Because of an
antibias in the culled halo distribution relative to mass, $\beta$ is
about $\sim 10$\% above unity on translinear to linear scales with
weak scale dependence.  Given the great difference in small-scale
clustering statistics between this dataset and the total mass
distribution, it is significant that
equation~(\ref{eq:pfilt_explicit}) still gives a good parameterization
of the anisotropy in redshift-space power on scales as small as
1000~\kms.

\section{Dispersion-Corrected Sampling Functions}

With a working model for the anisotropy of redshift-space power in the
presence of linear flows and small-scale dispersion, we now focus on
measuring $\beta$ from the linear contribution.  The strategy is to
determined the small-scale $\sigma_v$ in a redshift-space density
field, remove the effects of the dispersion signal as modeled by
equation~(\ref{eq:pfilt_disp}), and then to estimate the anisotropy in
Fourier modes on scales $\gsim 10$~\hmpc. In an infinite or periodic
dataset one could simply remove the dispersion signal by a
mathematically trivial deconvolution. For example, the redshift-space
distribution might be transformed into the Fourier domain and the
result multiplied by the inverse square-root of $F_\disp$. However, in
practice the finite size of a survey complicates the calculation of
the Fourier transform especially at large wavelengths (e.g., Fisher et
al. 1993).  Here we circumvent the problem of the Fourier transform
and deconvolution of the density field by making use of sampling
functions which are tuned to be orthogonal to the dispersion
signal. We consider only the variances in such samples and hence we
extract linear flow information without directly deconvolving the
redshift-space density field.

A sampling function $s$ gives a sample value $S$ at some
randomly chosen origin by projection
onto a density field $\rho$
\beq\label{eq:sampval}
S = \int \! d^3x \,  s(\myvec{x}) \rho(\myvec{x})/\overline{\rho} \ ,
\eeq
where $\overline{\rho}$ is the mean density.  It is straightfoward to
show that the variance in a set of many samples taken at random
locations in a density field is related to the power
spectrum $P$ by
\beq\label{eq:sampvar}
\sigma^2 = 
\int \! d^3k P(\myvec{k}) | {s}(\myvec{k}) |^2 \ .
\eeq
Of course in practical applications the power is inferred from
a discrete galaxy distribution which contains shot noise.  In this
case, equation~(\ref{eq:sampvar}) must be corrected by subtracting
the shot noise contribution,
\beq\label{eq:shotnoise}
\frac{1}{\overline{n}} \int \! d^3k | {s}(\myvec{k}) |^2 \ ,
\eeq
where $\overline{n}$ is the mean density of galaxies.

Anticipating the case when $P$ in equation~(\ref{eq:sampvar}) is set
to $\ps$, we choose $s$ to have a form such that in the Fourier domain
\beq\label{eq:sampsigv}
s(\myvec{k}) = \frac{s_\linflow(k,\mu)}{\sqrt{F_\disp(k,\mu)}}
\eeq
where $s_\linflow$ couples only to the signal from linear flows. We
assume that the samples are taken at a large distance from the
observer so that $\mu$ is constant. Clearly, the effect of the
denominator in equation~(\ref{eq:sampsigv}) is to perform the
deconvolution of dispersion signal at the level of random sampling.

We emphasize that in the method discussed here we evaluate only sample
variances; we never directly perform a deconvolution of the
redshift-space density field to remove shot noise and the effects of
small-scale dispersion.  Yet the filter given in equation
(\ref{eq:sampsigv}) might be used to do just that.  Indeed,
deconvolution with this filter and the inverse of the linear flow
filter in equation (\ref{eq:pfilt_linflow}) would enable the
real-space density field to be reconstructed from redshift-space data
even on linear scales which are affected by small-scale
dispersion. This would advance known reconstruction techniques (e.g.,
Yahil et al.~1994; Nusser \& Davis 1995; Tegmark \& Bromley 1996) by
introducing some correction for nonlinear dispersion.  However, the
filtering must be done with great attention to noise and finite
boundary effects.  For example, optimal (Wiener) filtering is
necessary to avoid the pitfalls of deconvolution in the presence of
noise (cf. Press et al. 1992, section 13.3).

With the nonlinear effects removed, the remaining problem is to
determine a measure of $\beta$ from the linear flow signal in the
sample variances. Bromley (1994) proposed an estimator that follows
from taking the derivative of some radially symmetric function $R$
along the direction of a unit vector $\myunitvec{n}$ at fixed
orientation relative to the distant observer:
\beq\label{eq:samprad}
s_\linflow(\myvec{x}) = \myunitvec{n}\cdot\nabla R(x) 
\ \ \ {\rm and} \ \ \ 
s_\linflow(\myvec{k}) = i \myunitvec{n}\cdot\myvec{k} R(k) \ .
\eeq
Here $R$ is specified to have the form of a 1D Gaussian,
\beq\label{eq:radfn}
R(x) = \exp(-x^2/2\lambda^2) \ ,
\eeq
where $\lambda$ defines the scale of the sampling function.
The effects of linear flows in redshift space can be isolated by
considering two different sampling functions constructed from
equations~(\ref{eq:sampsigv}) through (\ref{eq:radfn}), with
$\myunitvec{n}$ directed parallel and perpendicular to the observer's
line of sight, respectively.  The ratio of variances
$\sigma_\parallel$ and $\sigma_\perp$ from these two sampling
functions is easily related to $\beta$ using
equation~(\ref{eq:sampvar}) and the redshift-space power in
equation~(\ref{eq:pfilt_explicit}):
\beq\label{eq:omrat}
{\cal Q}(f) \equiv \frac{\sigma^2_{\parallel}}{\sigma^2_\perp} =
\frac{1 + \frac{6}{5}\beta + \frac{3}{7}\beta^2}
{1 + \frac{2}{5}\beta + \frac{3}{35}\beta^2} \ .  
\eeq
The desired measure of $\beta$ comes from inverting this expression.
Note that the result does not require any knowledge the form of the
real-space power spectrum as all dependence on $\pr$ cancels out in
the ratio $\cal Q$.

We use the numerical data described in \S2 to determine how well the
ratio of sample variances in equation~(\ref{eq:omrat}) is able to
recover the correct value of $\beta$.  Figure~2 illustrates the
recovery from redshift space data over a range of sample function
scales $\lambda$ using known values of the velocity dispersion $\sigv$
on megaparsec scales from the three simulation catalogs.  The method
works well for the mass particles and the full halo catalog; in both
cases the estimate of $\beta$ was within 10\% of its true value.  In
the culled halo subset the recovered $\beta$ was consistently about
10\% below the expected $\beta$ value of $\sim 1.1$ when the
$\sigv$ value of $\sim 384$~\kms\ at 1~\hmpc\ pair separation was used
in the sample functions. However, the number of halo pairs at a
megaparsec is separation is relatively small, and with $\sigv$ itself
rising fairly rapidly with separation at this scale, it is not
surprising that a higher value of $\sigv$, 450~\kms, gives a better
estimate of $\beta$.

Figure~2 shows strong promise for determination of $\beta$ with the
ratio-of-variances method using tuned sample functions. The figure
also illustrates the importance of taking into account small-scale
velocity dispersion. Even on redshift-space scales approaching three
times the pairwise velocity dispersion, we find that the recovered
$\beta$ is generally more than 50\% below its true value.

\begin{figure}[t]
 \centerline{
  \epsfxsize=3.0in\epsfbox{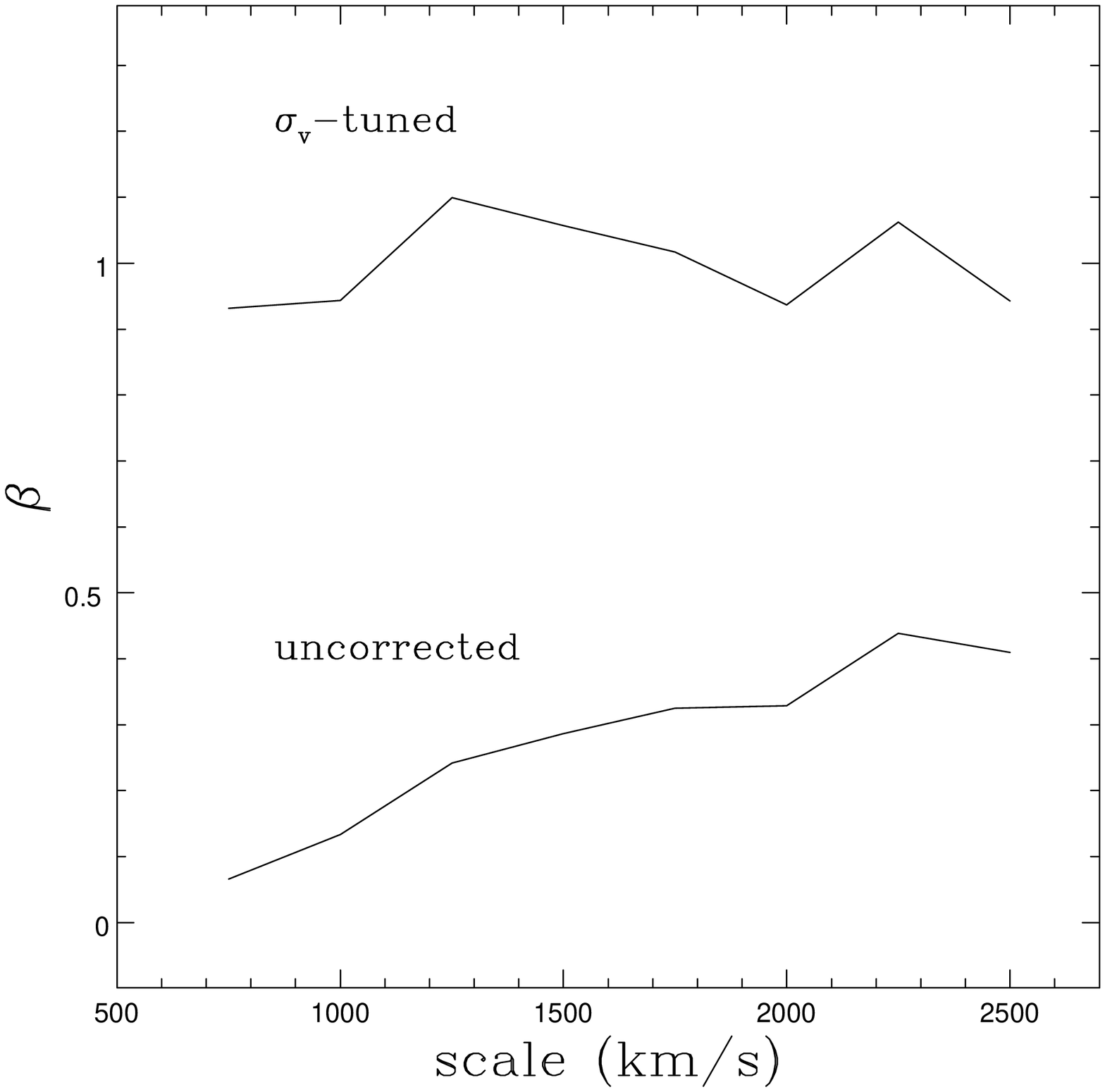}
  \epsfxsize=3.0in\epsfbox{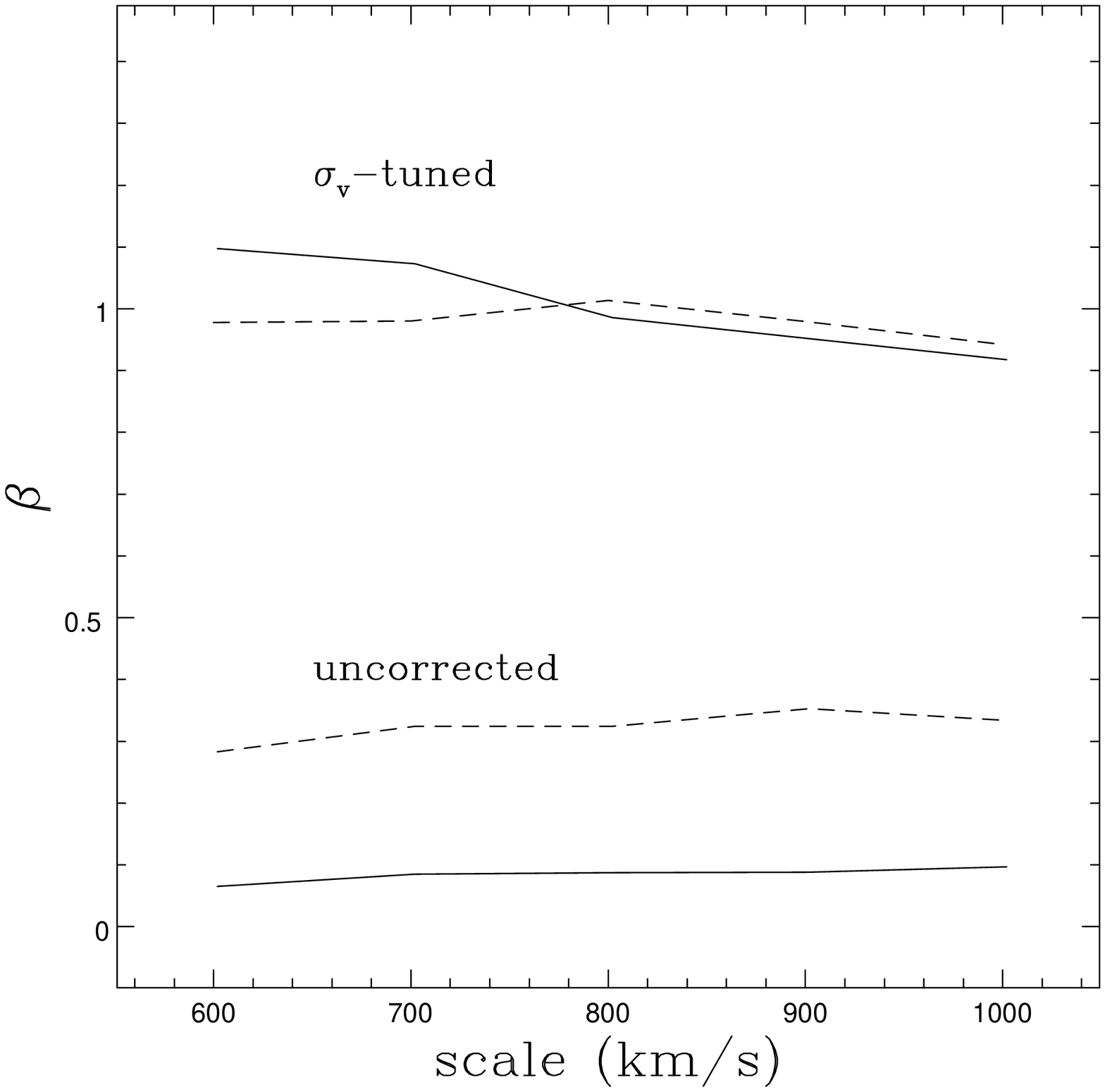}
}
\caption{\protect\small
The recovered $\beta$ from the ratio-of-sample-variances method.  In
the plots the upper curves were obtained using a $\sigv$ of
1,000~\kms, 850~\kms, and 425~\kms, respectively, and the lower curves
have no correction for dispersion ($\sigv = 0$). The results for the
mass distribution are shown in (a) while (b) contains the results for
the full halo catalog (solid curves) and the culled halo subset (dashed
curves).
}
\end{figure}

\section{The IRAS 1.2~Jy Survey}

We now apply the sampling method to the \iras\ redshift catalog.  This
is a flux-limited sample of 5,665 galaxies with sky coverage
of 87.6\% (Strauss et al. 1990; Yahil et al. 1991).  The flux limit in
the 60$\mu$m waveband causes the density of galaxies in redshift space
to fall off rapidly with distance; the number of galaxies in radial
bins peaks at $\sim$5,000~\kms, and the selection function $\phi$
which describes the fall-off has a value of less than 0.01 at
10,000~\kms. We take this latter distance to be the characteristic size
of this survey.

The sampling procedure used to generate $\cal Q$ from the IRAS
galaxies is more complicated than in the case of the pristine
numerical data.  First, each galaxy must be weighted by $1/\phi$ to
account for low-luminosity neighbors which were missed because of the
flux limit. Here we use a selection function of the form given by
Yahil et al. (1991, eq~[11] therein) with $a = 0.51$, $b = 1.84$, and
$hr_\ast = 5,440$~\kms.

The second complication in the analysis of the IRAS galaxies is that
shot noise is significant on scales below $\sim$1,000~\kms.  If
uncorrected it causes the sampling method to underestimate $\cal Q$
and $\beta$.  Here, we remove the shot noise contribution to the
sample variance by equation~(\ref{eq:shotnoise}); its amplitude for
each level of sampling scale is determined from artificial catalogs
generated by a Monte Carlo method in accordance with the selection
function of the IRAS galaxies.

A third complication is incomplete sky coverage.  So that we may
sample with the tuned functions without the limitations of avoiding
unobserved sky regions, we follow the strategy of Fisher, Scharf \&
Lahav (1994) who interpolated the IRAS galaxy density field across the
zone of avoidance and other unsampled regions. The interpolation
method is described in detail by Yahil et al.~(1991) as part of a
broader algorithm to reconstruct the real-space density field from the
IRAS data in redshift space.  Here, we include 705 artificial galaxies
in real space kindly provided by M.~Davis.  Since the artificial
galaxies represent a real-space density field, reconstructed under the
assumption that $\beta$ is unity, the line-of-sight mode amplitudes
are guaranteed to be smaller than in the redshift-space IRAS
catalog. This effect, along with the fact that any discontinuity with a
boundary running along the observer's line of sight will only add
power to wavemodes in the plane of the sky, can drive the ratio $\cal Q$
to systematically lower values, causing $\beta$ to be underestimated.


There is one final difference between the analyses of the numerical
data and of real galaxies. In the simulations the pairwise velocity
dispersion $\sigv$ at megaparsec scales, the essential ingredient of
the dispersion filter (eq,~[\ref{eq:pfilt_disp}]), is known precisely
from the 6D phase space distribution of particles or halos. In
contrast, $\sigv$ must be gleaned from redshift-space data in the IRAS
catalog, a procedure which introduces some uncertainty into $\sigv$.
Published values (e.g., Fisher et al. 1993) and our own estimate based
on the method of Davis \& Peebles (1983) suggest that $\sigv$ for the
\iras\ sample is 320~\kms\ with 1-$\sigma$ errors of $\sim 40$~\kms.
If the IRAS galaxies are unbiased tracers of mass, then the numerical
data suggest that this value should be appropriate for use in the
dispersion filter (eq.~[\ref{eq:pfilt_disp}]). If the IRAS population
is antibiased with respect to mass, as has been conjectured on the
basis of comparison with optical surveys, then the use of this value
may generate an underestimate of $\beta$. This conclusion is made on
the basis of similarity between our culled halo catalog and the IRAS
galaxies which both have relatively low number densities in cluster
cores (as compared to the full halo catalog or optically selected
galaxies) and rapidly rising $\sigv$ values on megaparsec scales.

After implementing the above modifications to the sampling procedure,
we analyzed the IRAS galaxies which lie inside a thick spherical shell
of redshift space with an inner radius of 500~\kms\ and an outer
radius chosen between 10,000--12,000~\kms.  We estimate uncertainty by
determining an approximate number of spatially independent samples
that can be measured with a given sampling function and make repeated
measurements of $\beta$ with that number of samples.  In Figure~3 we
show the recovered $\beta$ from the IRAS galaxies as a function of
sampling scale. In general, the recovered value has a lower bound at
the 1-$\sigma$ level which is above 0.4. The mean of the recovered
$\beta$ on scales of 600--1,500~\kms\ is in the range of 0.6--0.9,
depending on the outer radius of the sampled region. This sensitivity
shows up mostly at the larger sampling scales where we must balance a
desire to use as big a volume as possible against the increasing
sparsity of galaxies at large distances. The most stable estimate
comes from sampling scales around 1,000~\kms\ for which we find that
$\beta = 0.8^{+0.4}_{-0.3}$ at the $1-\sigma$ level.  As noted
above, the systematic error from zone-of-avoidance interpolation is
likely to cause this value of $\beta$ to be a slight underestimate of
the true value.  The similarity between the IRAS galaxies and our
culled halo catalog in \S{3} mentioned above suggests that our choice
of $\sigv = 320$~\kms\ may be too small, leading to a more significant
underestimate of $\beta$.  Indeed, $\sigv$ in the IRAS data climbs to
a value of 370~\kms\ at 2~\hmpc which leads to a $\sim$10\% increase
in $\beta$ over the estimates shown in Figure~3.

\begin{figure}[t]
 \centerline{
  \epsfxsize=4.0in\epsfbox{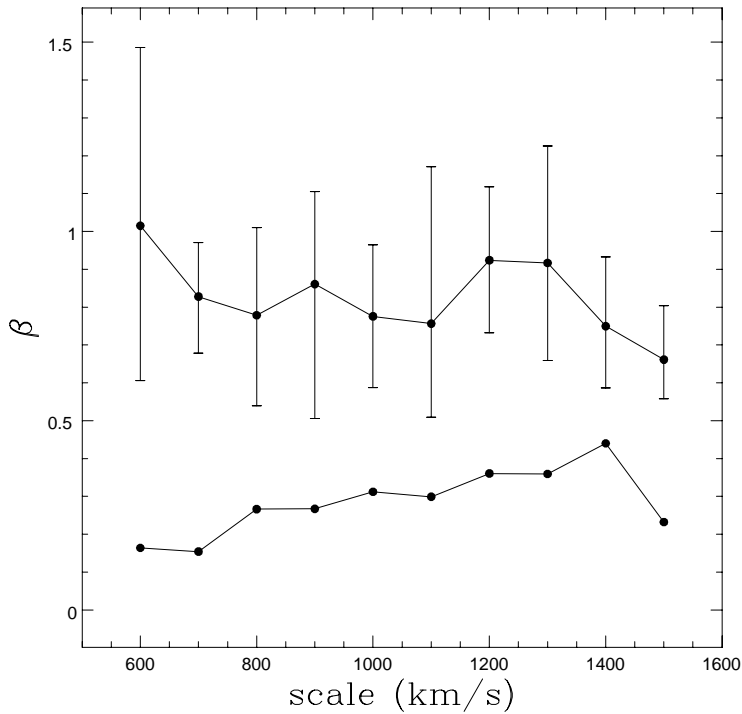}
}
\caption{\protect\small
The recovered $\beta$ from the \iras\ survey. The upper
curve comes from sampling with functions tuned to $\sigv = 320$~kms,
and the lower curve corresponds to zero dispersion.
The error bars show the 1-$\sigma$ uncertainties. 
}
\end{figure}

\section{Conclusions}

Here we have modeled the anisotropy in the redshift-space power
spectrum on the basis of Peacock \& Dodds' (1994) theoretical work and
our own high-resolution numerical simulations. We demonstrated that
the effects of nonlinear clustering on the power spectrum of both mass
and galaxy halos can be described by a simple multiplicative filter
function. The filter-based model for the anisotropy is seen to hold on
linear and translinear scales which are greater than the pairwise
velocity dispersion $\sigv$ at megaparsec. Having established the
domain of validity, we then proposed a redshift-space statistic, the
ratio of sample variances in equation~(\ref{eq:omrat}), which is both
impervious to the effects of nonlinear clustering and sensitive to
linear flows and the linear growth parameter $\beta$.

This work should serve as a cautionary reminder of the extent to which
the nonlinear distortions contaminate signal from linear flows (e.g.,
Fig.~2). In a measurement of power, the magnitude of contamination can
be determined from equation~(\ref{eq:pfilt_disp}); for example the
nonlinear signal reduces the power along the line of sight by $\sim
40$\% at scales $2\pi/k$ equal to five times $\sigv$.  In the case of
the \iras\ survey where $\sigv$ is $\sim 300$~\kms, the presence of
nonlinear distortion is important even at scales of 1,500~\kms.  For
optical surveys where $\sigv \sim 800$~\kms\ (e.g., Marzke et
al. 1995), the extent of contamination exceeds $4,000$~\kms.

The work presented here also gives the encouraging message that
nonlinear clustering does not prevent the accurate extraction of
$\beta$ from linear flows in redshift data. Figure~2 provides the
evidence based upon cosmological simulations of the greatest dynamical
range published to date.

The analysis of the \iras\ survey is intended to show that the method
introduced here gives reasonable results. Our estimate of $\beta =
0.8^{+0.4}_{-0.3}$ is consistent at the 2-$\sigma$ level with
virtually every published value based on an analysis of the IRAS
galaxies (e.g., Dekel et al. 1993; Fisher, Scharf \& Lahav 1994; Cole,
Fisher \& Weinberg 1995; Hamilton 1995), although it is higher than
most redshift-space measurements based exclusively on linear theory.
Our procedure is most similar in spirit to that of Hamilton (1995) who
obtained $\beta = 0.69^{+0.21}_{-0.19}$ using a merged \iras\ and the
optically selected QDOT catalog. Hamilton, working with harmonics of a
smoothed power spectrum, modeled the nonlinear noise in the same way
as we did, with an isotropic exponential distribution for pairwise
velocities. We hope that we have provided convincing evidence in \S2
that such a model is indeed robust.

We note that our estimate of $\beta$ from the \iras\ sample is
consistent with dynamical analyses that make use of real-space
information as well as redshift data. For example, Dekel et al. (1993)
find $\beta = 1.28 \pm 0.3$ on scales of 1,500--4,000~\kms. While
knowledge of both real-space and redshift distributions can yield a
more accurate measurement of $\beta$, the real-space data are
difficult to measure and therein lies the value in extracting $\beta$
directly from redshift space. Fisher et al. (1994) performed a pure
redshift-space analysis of the \iras\ galaxies by modeling $\xis$  and
obtained $\beta = 0.45^{+0.27}_{-0.18}$ on scales of 1,000--1,500~\kms,
well below the dynamical estimates at larger scales.  However, the
modeling of $\xis$ and its dependence on the uncertain behavior of
the velocity distribution function is difficult. In contrast, our
measure of $\beta$, based instead on a much simpler model in the
Fourier domain, is consistent with the large-scale dynamical results.

The ultimate hope is that our measure may be applied to forthcoming
optical surveys which exhibit high $\sigv$ values (e.g., Marzke et
al. 1995). With the ability to accurately determine $\beta$ down to
translinear scales in spite of nonlinear clustering effects, our
measure can provide high statistical significance in a correlation
analysis of redshift-space data.

\acknowledgements

We are grateful to Peter Quinn and the Mount Stromlo Observatory for
hospitality at the Heron Island Workshop on Peculiar Velocities.  This
work was supported under the auspices of the U.S. Department of Energy
and by the NASA HPCC program. The Cray Supercomputer used in this
investigation was provided through funding from the NASA Offices of
Space Sciences, Aeronautics, and Mission to Planet Earth.

\end{document}